\documentclass[3p,times,procedia]{elsarticle}
\usepackage{nupha_ecrc}
\usepackage[percent]{overpic}

\volume{00}

\firstpage{1}

\journalname{Nuclear Physics A}

\runauth{P.~M.~Jacobs, A. Schmah, et al. }

\jid{nupha}

\jnltitlelogo{Nuclear Physics A}

\usepackage{amssymb}

\usepackage{lineno}

\usepackage[figuresright]{rotating}

\newcommand{\CommentBlock}[1]{}

\newcommand{\AuAu}{Au+Au}
\newcommand{\PbPb}{Pb+Pb}

\newcommand{\pp}{pp}

\newcommand{\sqrtsNN}{\ensuremath{\sqrt{s_{\mathrm {NN}}}}}

\newcommand{\rr}{\ensuremath{R}}

\newcommand{\gev}{\ensuremath{\mathrm{GeV/}c}}

\newcommand{\antikT}{anti-\ensuremath{k_\mathrm{T}}}

\newcommand{\pT}{\ensuremath{p_\mathrm{T}}}

\newcommand{\pTjetch}{\ensuremath{p_\mathrm{T,jet}^\mathrm{ch}}}

\newcommand{\dNjetdpTdphi}{\ensuremath{\frac{\rm{d}^{2}N^\mathrm{AA}_{jet}}{\mathrm{d}\pTjetch\mathrm{d}\dphi}}}

\newcommand{\Ntrig}{\ensuremath{\mathrm{N}^\mathrm{AA}_{\rm{trig}}}}

\newcommand{\ICP}{\ensuremath{I_\mathrm{CP}}}

\newcommand{\pTrawi}{\ensuremath{p_\mathrm{T,jet}^\mathrm{raw,i}}}

\newcommand{\pTreco}{\ensuremath{p_\mathrm{T,jet}^\mathrm{reco,ch}}}

\newcommand{\pTrecoi}{\ensuremath{p_\mathrm{T,jet}^\mathrm{reco,i}}}

\newcommand{\rhoAi}{\ensuremath{\rho\cdot{A_\mathrm{jet}^\mathrm{i}}}}

\newcommand{\phiTrig}{\ensuremath{\varphi_\mathrm{trig}}}
\newcommand{\phiJet}{\ensuremath{\varphi_\mathrm{jet}}}
\newcommand{\dphi}{\ensuremath{\Delta\varphi}}

\newcommand{\AAtohjet}{\mathrm{AA}\rightarrow\rm{h}+{jet}+X}
\newcommand{\AAtoh}{\mathrm{AA}\rightarrow\rm{h}+X}

\usepackage{color}

\begin{document}

\begin{frontmatter}



\dochead{}

\title{Measurements of jet quenching with semi-inclusive charged jet distributions in \AuAu\ collisions at \sqrtsNN=200 GeV}


\author{P.~M.~Jacobs\corref{cor1} and A.~Schmah for the STAR Collaboration}

\address{Lawrence Berkeley National Laboratory, Berkeley CA, USA}

\cortext[cor1]{pmjacobs@lbl.gov,aschmah@lbl.gov}

\begin{abstract}
We report measurements of jet quenching in \AuAu\ collisions at \sqrtsNN=200 GeV, based on the semi-inclusive distribution of reconstructed charged particle jets recoiling from a high \pT\ hadron trigger. Jets are reconstructed with the \antikT\ algorithm (\rr=0.2 to 0.5), with low IR-cutoff of track constituents ($\pT>0.2$ \gev). Uncorrelated background is corrected using a novel mixed-event technique, with no fragmentation bias imposed by the correction procedure on the accepted recoil jet population. Corrected recoil jet distributions, reported in the range $0<\pTjetch<30$ \gev, are used to measure jet yield suppression, jet energy loss, and intra-jet broadening. The first search for QCD Moli{\`e}re scattering of jets in hot QCD matter at RHIC is reported.
\end{abstract}


\end{frontmatter}



\vspace{0.5 cm}

The interaction of energetic jets with hot QCD matter (``jet quenching'') 
provides unique probes of the Quark-Gluon Plasma (QGP) generated in high energy
collisions of heavy nuclei. Comprehensive understanding of jet quenching 
requires measurements of reconstructed jets and their correlations. Measurement of
reconstructed jets in the  high-multiplicity environment of heavy 
ion collisions is challenging, however, because of the large and
inhomogeneous backgrounds in such events. In this proceedings the STAR 
Collaboration at RHIC reports measurements of jet 
quenching in central and peripheral \AuAu\ collisions at \sqrtsNN=200 GeV, using 
an observable designed to address this challenge: the semi-inclusive distribution 
of reconstructed charged particle jets recoiling from a high \pT\ trigger
hadron. A similar measurement has been carried out by the ALICE Collaboration at the LHC, for \PbPb\ collisions at \sqrtsNN=2.76 TeV \cite{Adam:2015doa}. 

The data were recorded by STAR during the 2011 RHIC run with \AuAu\ collisions 
at \sqrtsNN=200 GeV, using a minimum bias trigger. Offline analysis is carried 
out using charged tracks measured by the STAR Time Projection Chamber (TPC). 
Events are classified in percentile intervals of uncorrected multiplicity of 
charged tracks within $|\eta|$ $<$0.5; this analysis uses events in the 
0--10\% (``central'') and 60\%--80\% (``peripheral'') percentile intervals. One 
trigger hadron per event is selected randomly from all observed charged 
particles with $\pT>9$ \gev. Charged jets, composed of charged tracks, are 
reconstructed using the \antikT\ algorithm \cite{FastJetAntikt} with the 
boost-invariant \pT-recombination scheme, for
resolution parameter \rr\ = 0.2, 0.3, 0.4 and 0.5. The raw reconstructed jet 
energy for each jet candidate $i$,
\pTrawi, is shifted by the estimated background
energy in the jet area,

\begin{equation}
\pTrecoi=\pTrawi - \rhoAi.
\label{eq:pTraw}
\end{equation}

\noindent
where $\rho$ is the estimated median background energy density in the event, and 
$A_i$ is the jet area. Jet acceptance is $|\eta_{\rm{jet}}|<1.0-\rr$, based on the jet 
centroid. A jet area cut  suppresses background jets while preserving  high efficiency for true hard jets. Since $\rho$ is the median background level in the event, \pTrecoi\ 
can be negative; we expect that region to be dominated by background that is 
uncorrelated to the trigger. 
 
The semi-inclusive recoil jet distribution is 
equivalent to the ratio of inclusive production cross sections,

\begin{equation}
\frac{1}{\Ntrig}\cdot\dNjetdpTdphi= \left(
\frac{1}{\sigma^{\AAtoh}} \cdot
\frac{\rm{d}^2\sigma^{\AAtohjet}}{\mathrm{d}\pTjetch\mathrm{d}\dphi}\right),
\label{eq:hJetDefinition}
\end{equation}

\noindent
where AA denotes \pp\ or \AuAu\ collisions; $\dphi=|\phiTrig - \phiJet|$; 
$\sigma^{\AAtoh}$ is the inclusive
cross section to generate a trigger hadron; 
 and 
$\rm{d}^2\sigma^{\AAtohjet}/\rm{d}\pTjetch\mathrm{d}\dphi$ is the
incluisve differential cross section for coincidence production of a trigger
hadron and recoil jet. The recoil acceptance is $|\pi-\dphi|<\pi/4$. After correction for uncorrelated background, the distribution in Eq. 
\ref{eq:hJetDefinition} is the
absolutely normalized recoil jet 
yield correlated with a single high-$Q^2$ interaction in the \AuAu\ collision. 

The distribution of uncorrelated background jets is determined by a mixed 
event (ME) procedure, in which events are mixed in exclusive bins of multiplicity, 
primary vertex $z$, and event-plane orientation. The full analysis, including 
jet reconstruction, is then rerun on the ME events. 

\begin{figure}[htbp] 
\centering
\includegraphics[width=0.4\textwidth]{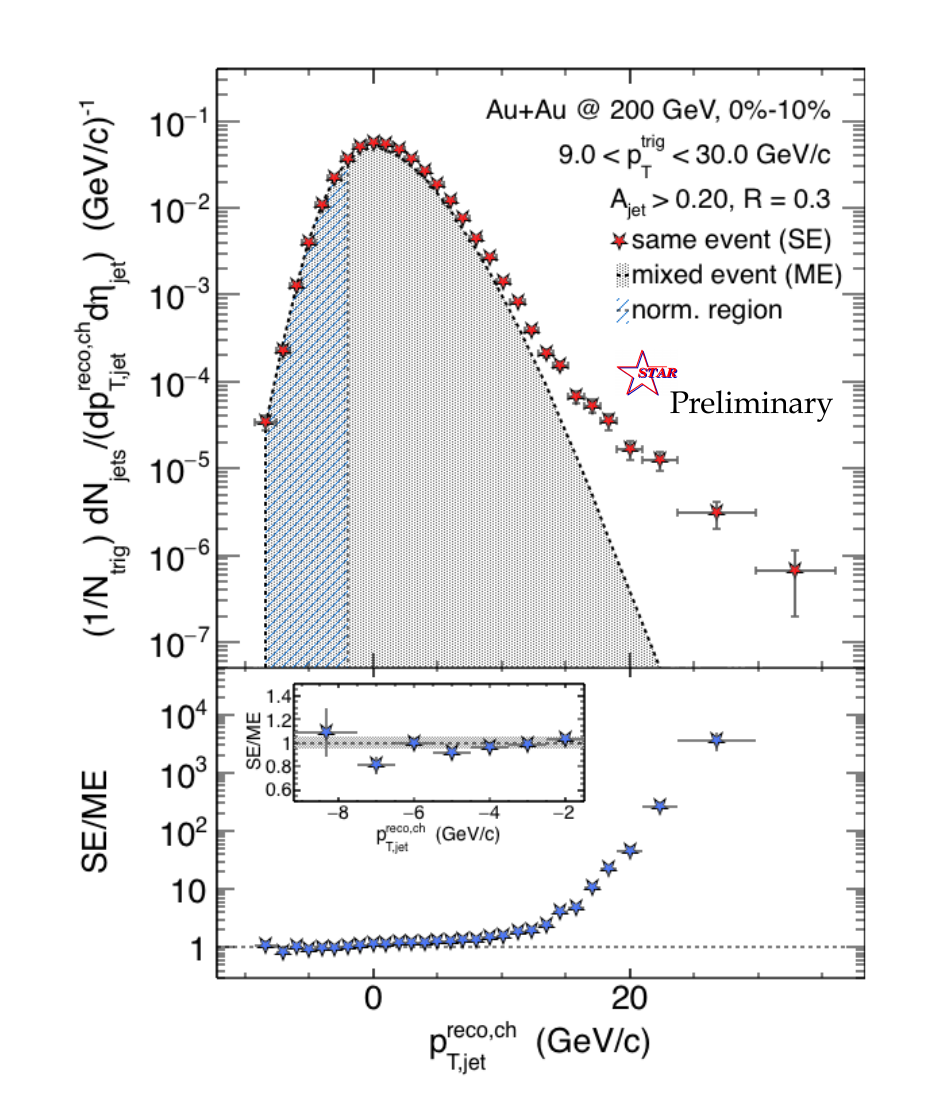}
\begin{overpic}
[width=0.4\textwidth]{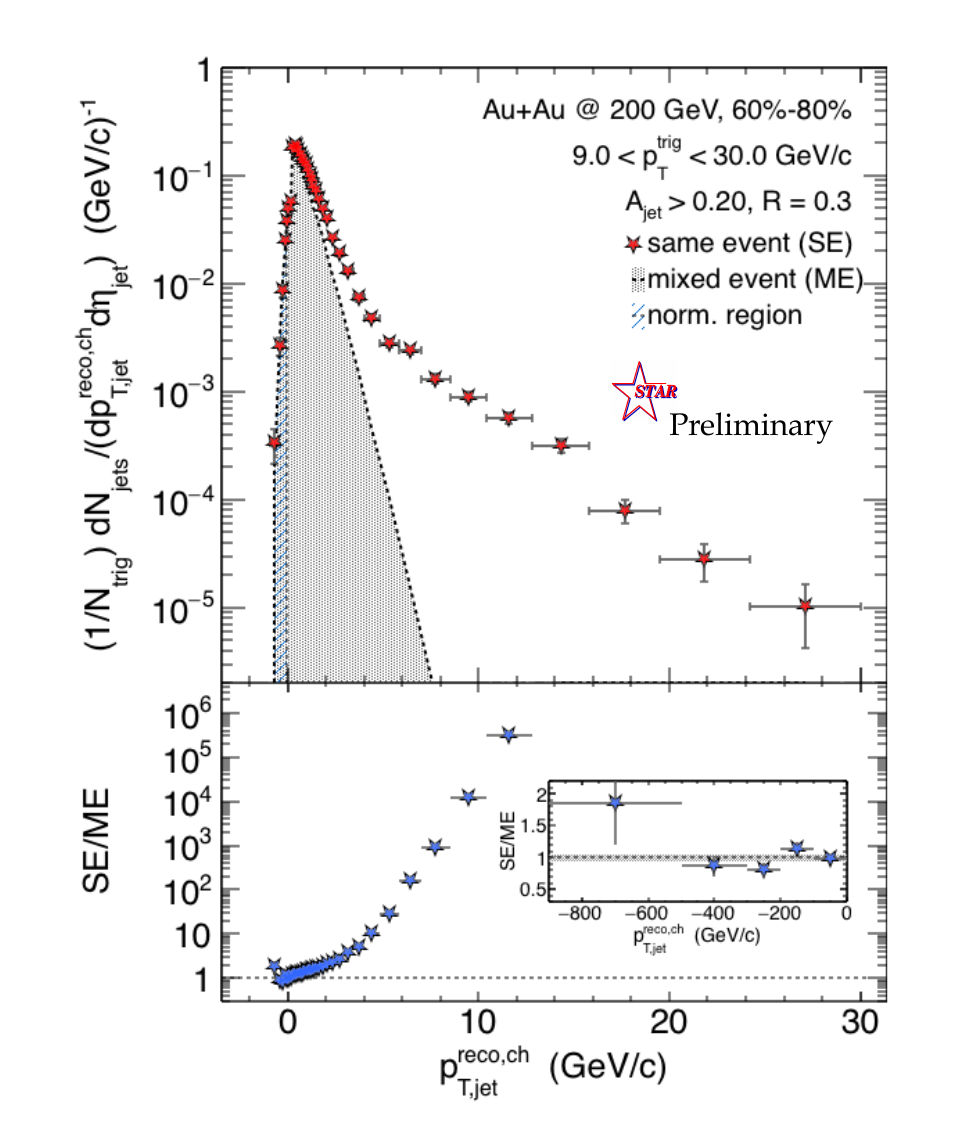}
\put(68,16){\tiny  $\times 10^{-3}$}
\end{overpic}
\caption{Upper panels: Uncorrected recoil jet distributions for \rr=0.3 in central (left) and peripheral (right) \AuAu\ collisions at \sqrtsNN=200 GeV, from both real (SE, red points) and ME events (shaded region). Lower panels: ratio of distributions SE/ME.}
\label{fig:RawData}
\end{figure}

Fig. \ref{fig:RawData} shows
uncorrected recoil jet distributions in real (SE) and ME events, for central and peripheral collisions. The ME jet distributions describe the real event distributions accurately for $\pTreco<0$, where uncorrelated background is expected to dominate the jet yield. At large positive \pTreco, the SE yield is much larger than the ME yield, as expected if the correlated yield dominates. The distribution of correlated recoil jet yield is determined by subtracting the ME from the SE distribution. This raw correlated distribution is then corrected by an unfolding procedure for instrumental effects and \pT-smearing due background. The corrected recoil distributions are a function of \pTjetch, the corrected charged-jet \pT.

\begin{figure}[htbp] 
\centering
\includegraphics[width=0.4\textwidth]{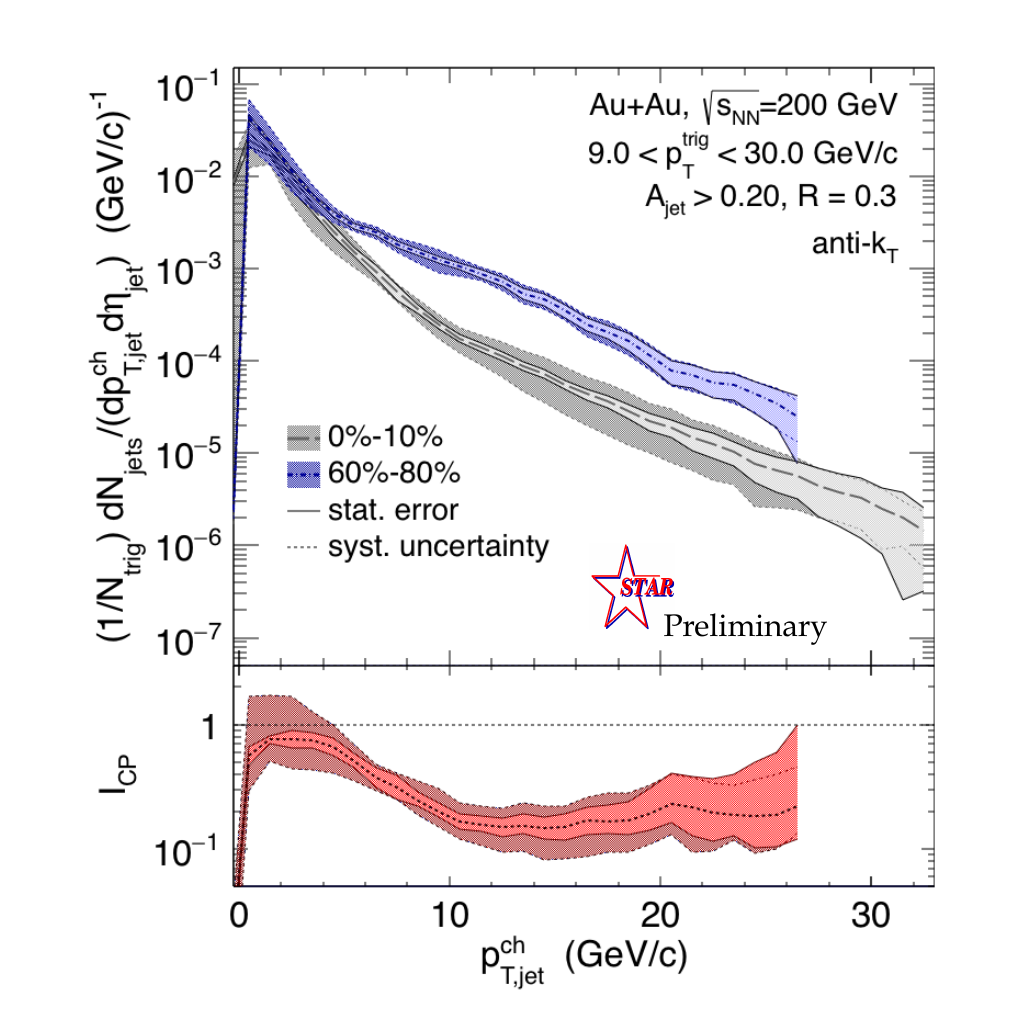}
\includegraphics[width=0.4\textwidth]{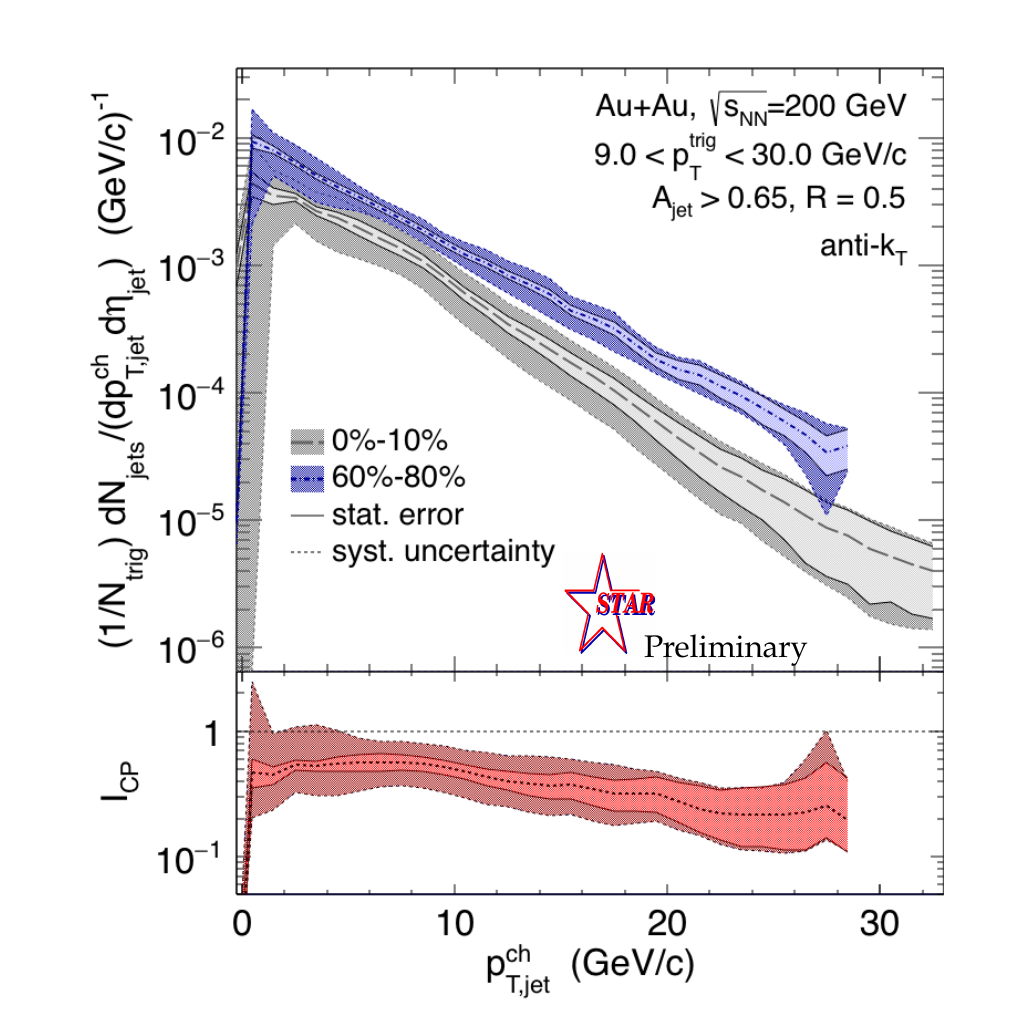}
\caption{Upper panels: Corrected recoil jet distributions for peripheral and central \AuAu\ collisions, for \rr=0.3 (left) and 0.5 (right). Lower panels: \ICP, the ratio of central to peripheral yield.}
\label{fig:ICP}
\end{figure}

Figure \ref{fig:ICP}, upper panels, show the corrected semi-inclusive 
recoil jet distributions for peripheral and central \AuAu\ collisions, for 
\rr=0.3 and 0.5. The distributions are shown for corrected 
$\pTjetch>0$, with contribution from all recoil jets. The lower panels show 
\ICP, the ratio of the central to peripheral 
distributions. For $\pTjetch>10$ \gev, there is significant yield suppression in 
central collisions for \rr=0.3, with less suppression for \rr=0.5. In a 
range where the ratio is flat, the suppression in \ICP\ can 
be expressed equivalently as a horizontal shift between the distributions. In 
the range $10<\pTjetch<20$ \gev, the shift is $-6.3\pm0.6\pm0.8$ \gev\ for 
\rr=0.3 and $-3.8\pm0.5\pm1.8$ \gev\ for \rr=0.5. ALICE has made a similar 
measurement for \PbPb\ collisions at \sqrtsNN=2.76 TeV, finding a shift in the 
recoil jet distribution between \pp\ and central \PbPb\ collisions to be 
$-8\pm2$ 
\gev\ in the range $60<\pTjetch<100$ \gev\ \cite{Adam:2015doa}. This shift may 
be 
interpreted as energy transport out of the jet cone due to jet quenching, in 
other words a direct measurement of partonic energy loss \cite{Adam:2015doa}. 
These measurements provide the first quantitative comparison of 
reconstructed jet quenching at RHIC and LHC.

\begin{figure}[htbp] 
\centering
\begin{overpic}
[width=0.4\textwidth]{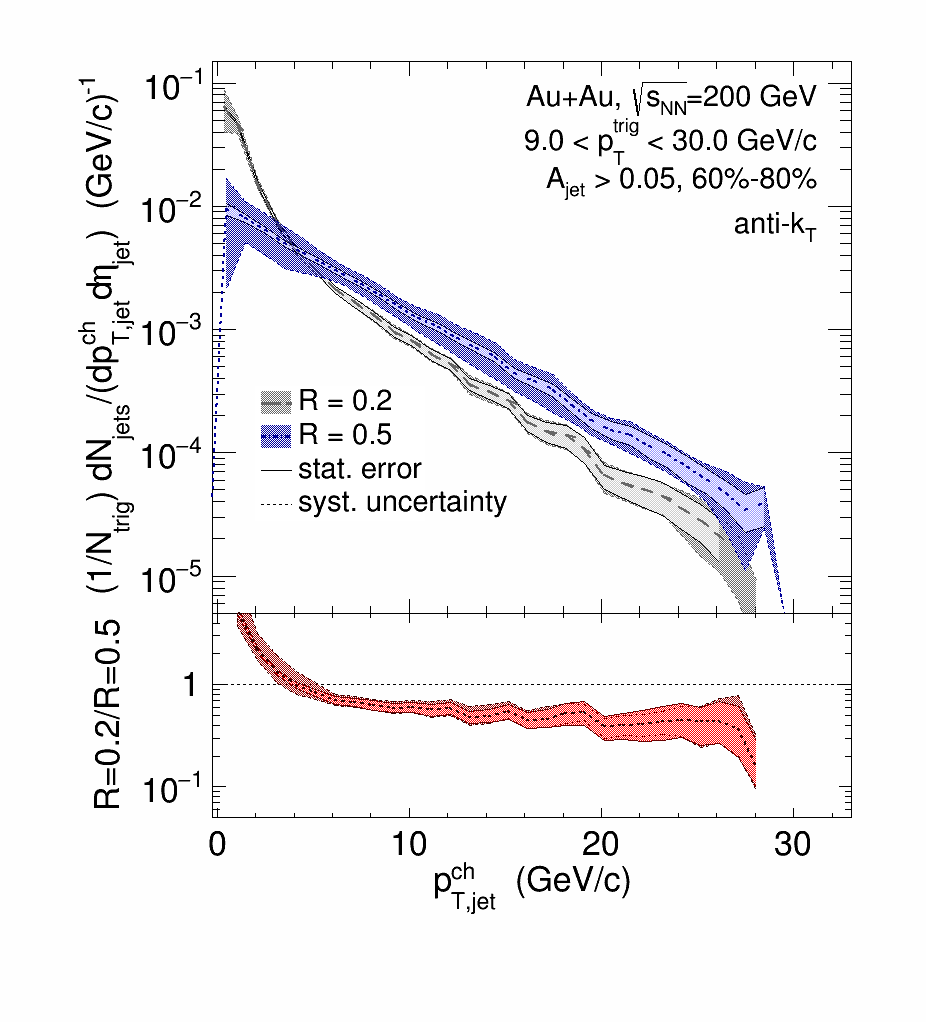}
\put(50,70){\footnotesize STAR preliminary}
\end{overpic}
\begin{overpic}
[width=0.4\textwidth]{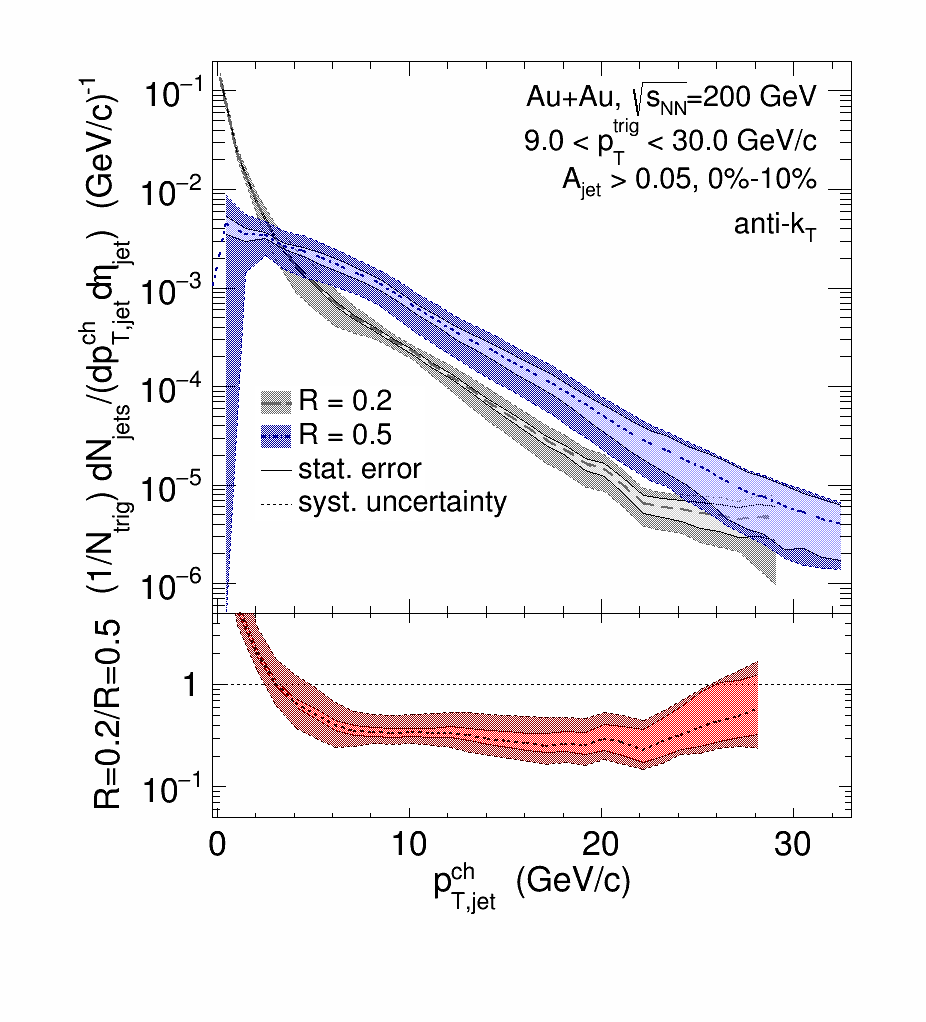}
\put(50,70){\footnotesize STAR preliminary}
\end{overpic}
\caption{Upper panels: corrected recoil distributions for \rr=0.2 and \rr=0.5, separately for peripheral (left) and central (right) \AuAu\ collisions. Lower panels: ratio of recoil yields for \rr=0.2/\rr=0.5.}
\label{fig:Ratio_2_5}
\end{figure}

For jets in \pp\ collisions, the ratio of inclusive cross sections or 
semi-inclusive yields at different \rr\ probes the distribution of intra-jet energy flow transverse to the jet axis
\cite{Soyez:2011np,Abelev:2013fn,Chatrchyan:2014gia,Adam:2015doa}. For jets in 
heavy ion collisions, such ratios provide experimentally robust observables of the 
modification of intra-jet structure due to quenching, \cite{Adam:2015doa}. Fig. 
\ref{fig:Ratio_2_5} shows the distribution of corrected 
recoil jet yields for \rr=0.2 and \rr=0.5 (upper panels) and their ratio (lower panels), 
for peripheral and central \AuAu\ collisions. In the region $\pTjetch>10$ \gev, the 
weight of the distribution is lower for central than for peripheral collisions, suggesting 
medium-induced intra-jet broadening within angle $\rr<0.5$. 
However, the current uncertainty bands are also consistent with 
absence of medium-induced broadening in this range. A similar picture is 
obtained from current measurents of \PbPb\ collisions at the LHC \cite{Adam:2015doa}.


\begin{figure}[htbp] 
\centering
\begin{overpic}
[bb = 0 0 184 208,clip, width=0.3\textwidth]{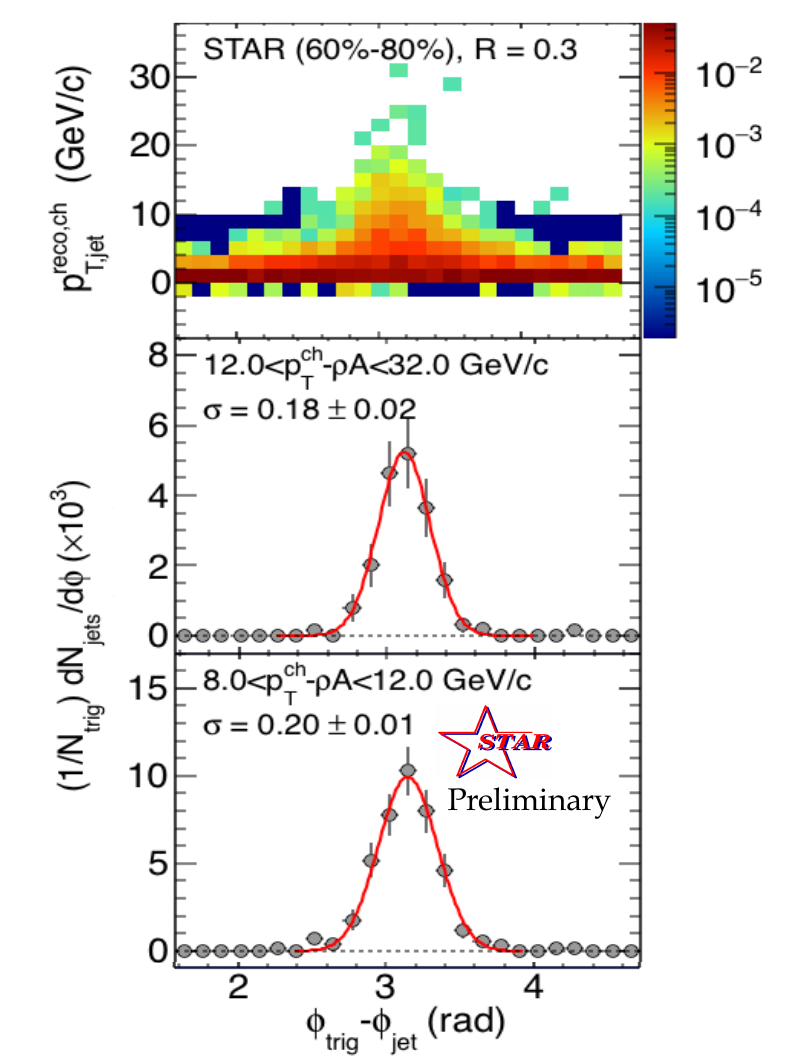}
\put(28,80){\footnotesize  60\%-80\%}
\end{overpic}
\begin{overpic}
[bb = 0 0 184 208,clip, width=0.3\textwidth]{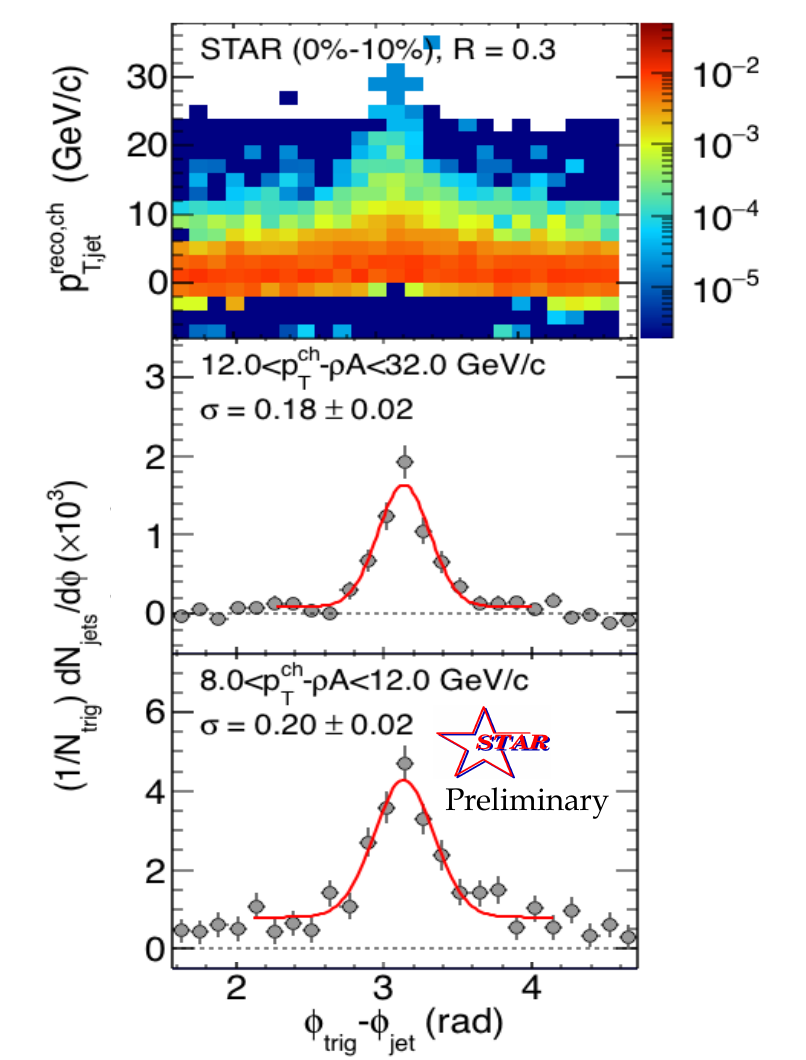}
\put(28,80){\footnotesize  0-10\%}
\end{overpic}
\caption{Distribution of \dphi\ for two bins in uncorrected jet energy in peripheral (left) and central (right) \AuAu\ collisions at \sqrtsNN=200 GeV. Uncorrelated yield has been subtracted, but jet energy is not corrected for instrumental and background effects.}
\label{fig:dphi}
\end{figure}

Acoplanarity of a di-jet pair in vacuum arises from radiation emitted at angles 
greater than \rr\ to the jet centroid. In heavy ion collisions, medium-induced di-jet acoplanarity may 
also occur \cite{D'Eramo:2012jh,Wang:2013cia}. At sufficiently large angular deviation, hard 
scattering off quasi-particles in the medium (the QCD analog to Moli{\`e}re 
scattering) may dominate the angular distribution between trigger axis and 
recoil jet, with soft multiple scattering 
and radiative processes being sub-dominant \cite{D'Eramo:2012jh}. Such 
measurements could discriminate between a medium that has discrete 
quasi-particles, and one that is effectively continuous at the $Q^2$ scale being 
probed \cite{D'Eramo:2012jh}. 

Figure \ref{fig:dphi} shows the first search for QCD Moli{\`e}re scattering at 
RHIC. The \dphi\ distribution of recoil jets (Eq. \ref{eq:hJetDefinition}) is 
shown for two bins at low jet energy in peripheral and 
central collisions, for jets with \rr=0.3 . In this case, uncorrelated background is corrected by 
subtraction of the ME distribution, but the jet energy is not corrected for 
instrumental effects and background fluctuations. 

Since the distribution is 
semi-inclusive, and thereby absolutely normalized, the 
yield in the large-angle tails relative to the 
trigger axis ($|\pi-\dphi|>\sim1$) can provide a direct measurement of the rate 
of Moli{\`e}re scattering. Comparison of the tails of the distributions indeed 
shows non-zero yield at large angles for the 
lower jet energy bin in central collisions, similar to the signal expected from 
Moli{\`e}re scattering. 
However, other effects such as flow may contribute to the signal in this region, 
and we do not claim evidence of 
Moli{\`e}re scattering based on this distribution. It rather serves to 
illustrate 
both the potential and the subtleties of this measurement. The fully corrected 
distribution, and future measurements with 
higher integrated luminosity, 
will provide constraints on the Moli{\`e}re scattering process.





\bibliographystyle{elsarticle-num}
\bibliography{references_QM}







\end{document}